\documentclass[journal]{IEEEtran}
\usepackage{graphicx}
\begin{document}
\title{Network representation of cardiac interbeat intervals
for monitoring restitution of autonomic control  for heart transplant
patients. }

\author{Danuta Makowiec, Stanis\l aw Kryszewski, Beata Graff, Joanna
Wdowczyk-Szulc, Marta \.Zarczy\'nska- Buchnowiecka, Marcin Grucha\l a
and Andrzej Rynkiewicz
\thanks{Danuta Makowiec and Stanis\l aw Kryszewski  are  with
Gda\'nsk University, Poland,  e-mail: fizdm@univ.gda.pl.}
\thanks{Beata Graff, Joanna Wdowczyk-Szulc, Marta \.Zarczy\'nska-Buchnowiecka,
Marcin Gru\-cha\-\l a and Andrzej Rynkiewicz are with Gda\'nsk Medical
University, Poland. }
}

\markboth{Proceedings of the 7{\tiny th} ESGCO 2012, April 22-25, 2012,
Kazimierz Dolny, Poland}%
{Makowiec \MakeLowercase{\textit{et al.}}: Network representation of
cardiac interbeat intervals for monitoring restitution of autonomic control
for heart transplant
patients. }

\maketitle

\begin{abstract}
The aim is to present the ability of a network of transitions as a nonlinear 
tool providing a graphical representation of a time series. 
This representation is used for cardiac RR-intervals
in follow-up observation of changes  in heart rhythm of  patients
recovering after heart transplant.
\end{abstract}

\begin{IEEEkeywords}
heart transplant, heart rate variability, graphical representation
\end{IEEEkeywords}


\section{Introduction}
The patient's life is usually endangered before the decision to
transplant the heart is taken. On the other hand, in many cases it is amazing how
the patient's organism recovers after heart transplant (HTX)
\cite{Toledo2002}.

It is generally believed that the time intervals between subsequent
heart contractions (so-called RR-intervals) carry the information
about the cardiac control system mainly driven by autonomic nervous system \cite{TASK}. However, heart transplantation interrupts the possibility of autonomic control over the heart beating. Therefore,  heart rate variability (HRV)
in patients after HTX is low, regardless of the time elapsed since
surgery. It is controversial whether the cardiac reinnervation
occurs after HTX \cite{Toledo2002}. But it is expected that progressive
reinnervation sets in and it is a good prognosis for survival.

We believe that the changes that occur in  heart rhythm may provide the first signals of  the recovery of cardiac control. Furthermore, these signals should be connected with the increasing influence of the sympathetic nervous system. Therefore we hope that observing these changes, with the help of carefully selected tools, we can describe the process of cardiac reinnervation.

The considerable success of the network theory in various fields of
research (see, e.g., \cite{Donner, Costa}) motivated us to explore
these ideas in the analysis of HRV time series. Tools of the complex networks allow one to resolve important and complementary properties of a dynamical system. For example, it is possible to study spatial dependencies
between individual observations instead of temporal correlations.
In the following  we continue our earlier studies
(see \cite{Makowiec2011}) on  networks of transitions applied to study recordings of  time intervals between subsequent cardiac contractions in  patients after HTX.

We will also raise the question whether these transitions build a monotonic  sequence of accelerations or decelerations. We are of opinion that sequences of monotonic accelerations or decelerations may indicate response of  the cardiac system  to some special needs of the organism. Hence, these studies may have a chance to offer  additional insights into the emergence of the heart regulatory control.

It must be emphasized that all graphical representations of
the discussed networks are produced with the {\it Pajek}
software package \cite{pajek}.

\section{Methods}

\subsection{The patient group}

We analyze  24-hour sequences of 23 ECG signals comprising of the
intervals between two successive R waves of sinus rhythm.
These signals were taken from 11 patients recovering after heart
transplantation in the First Cardiology Clinic of Gda\'nsk Medical
University. The recordings were taken from the same patients within
different periods after surgery. Therefore, we have 3 recording from
two patients, 2 recordings  from 8 patients and 1 recording from
one patient. We considered signals taken from two weeks
to 38 months after HTX.

From each RR-signal we have carefully selected a sequence
of 15,500 points corresponding to nocturnal rest of a patient.
There are two reasons why we investigate the nocturnal heart rhythm.
The first one is that during the sleep the central nervous system
is less dependent on the patient's intentions, and therefore
we may have a more direct insight into reflexes regulating
the cardiac rhythm. Moreover, the nocturnal recordings appear
to be less perturbed by artifacts what enables us to study
sufficiently long and consistent signals.

To avoid influence of artifacts (errors in detecting the R-wave)
the consecutive RR-intervals were thoroughly reviewed.
The parts, which consist of at least 500 normal-to-normal intervals,
were identified. If two such parts  were separated by artifacts
or ectopic beats of the length smaller than 10, then the gap was
edited manually. To preserve time chronology, the corresponding
RR-intervals were interpolated by the value of median from the last
normal-to-normal seven events. Additionally, the value of  median
was confronted with the total length of the edited gap.

\subsection{Network representation of RR-signal}

In general, the construction of the transition network \cite{Donner} is based on  the concept of phase space. The phase space represents all possible values of  studied dynamical system partitioned into mutually disjoint sets.
Since the recorded values of RR-intervals have well-separated magnitudes
then partitioning of the value space of RR-intervals is natural. Assuming these values as vertices of a network, we  represent each pair  of consecutive in time RR-intervals as a transition between these vertices.

\begin{figure}[h]
\includegraphics[height=23em]{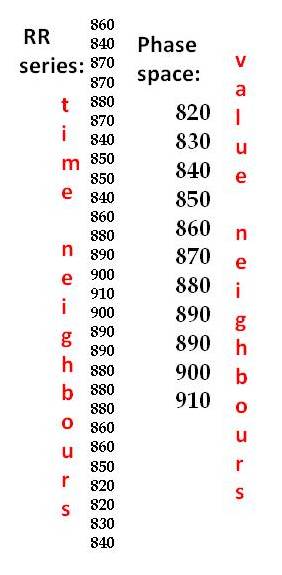}
\hfill
\includegraphics[height=19em]{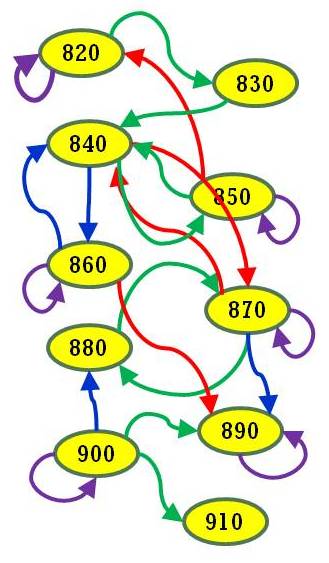}
\caption{Construction of a network from a time series. {\it Left}:
space values from  a time series discretized at $\Delta =10msec$.
{\it Right}:  a time series as a path on a network of values
of phase space.}
\label{Fig1}
\end{figure}

More precisely, the $N$ values of a recording with RR-intervals
 $ \{ RR_1, RR_2, \dots,RR_N \} $  are uniformly discretized (rounded)
 with accuracy $\Delta = 10$~ms.
Then, in order to determine the phase space, a set of ordered distinct values $RR^{MIN}\!=\!RR^{(1)}\! <\! RR^{(2)}\! < \! \dots \!  < \! RR^{(K)}
\! = \! RR^{MAX}$ is extracted.
These $K$ different values label different  vertices in the network.
This is depicted in Fig. \ref{Fig1} left where the first column gives
the sequence of RR intervals (already discretized), the second column
gives the ordered values from $RR^{MIN}$ to $RR^{MAX}$, thus
constituting the phase space of the  signal.
The right panel shows the studied time series as a transition
network. Here, the labels of vertices correspond to values from the phase space. Vertices are arranged from $RR^{MIN}$(top) to $RR^{MAX}$(bottom). An arrow between two vertices is plotted if the corresponding two RR-intervals represent a pair of the consecutive values in the time series.

It appears  that neighbors  in time are often also neighbors in values what results that the transition network takes the linear shape. But  to improve the visualization of the transition properties we propose a ladder  presentation  of the network.
Namely, vertices are placed alternatively in the left and right columns.

Moreover, to classify differences among the transitions we use the following coloring scheme:\\[2mm]
-- no change in the value,  i.e.,  $RR_i=RR_{i+1}$ : violet;\\
-- to a nearest neighbor, i.e., $|RR_i-RR_{i+1}|= \Delta $ : green;\\
-- to a next neighbor,  i.e.,  $|RR_i-RR_{i+1}|= 2\Delta $ : blue;\\
-- to a second neighbor, i.e., $|RR_i-RR_{i+1}|= 3\Delta $ : red;\\
-- to other neighbors, i.e., $|RR_i-RR_{i+1}|> 3\Delta $ : black. \\[2mm]
\begin{figure}
    \includegraphics[height=17.2em]{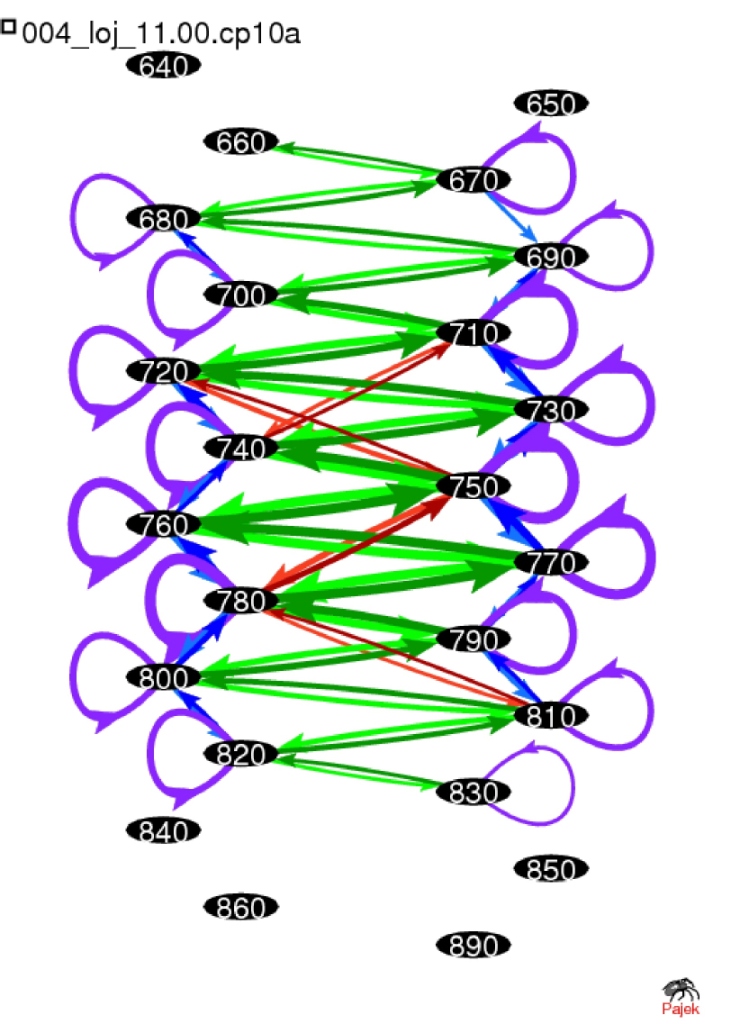}
    \includegraphics[height=17.2em]{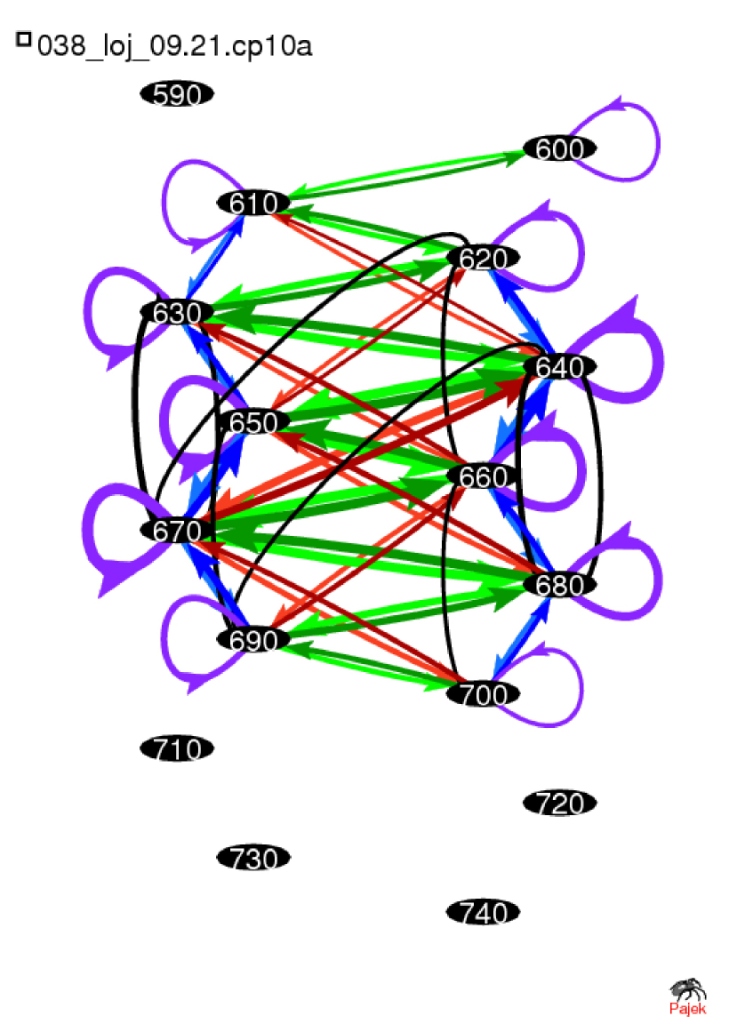}
\caption{Typical networks obtained for a patient (here, called
{\it loj} ) when the patient  was 4 (left) and 38 (right) months
after HTX. For better readability the transitions with probability
less than 0.1\% are omitted but all vertices which  correspond
to all recorded RR-intervals are plotted. The widths of arrows represent
logarithms of counts of the particular transitions.}
\label{Fig2}
\end{figure}

Fig.~\ref{Fig2} shows two transition networks obtained for a patient called {\em loj}.  The left network represents heart rhythm after 4 months elapsed after HTX, the rigth one 34 months later. The important message  of this example is that when the time after HTX increases the number of transitions other than to nearest neighbors also increases (in the right panel there are more red and black arrows). We believe that it is a good symptom.

It should be explained that widths of the transition arrows in
Fig.~\ref{Fig2} and all further network plots are determined by
logarithms of the frequencies of particular transitions.

Study of the transitions between RR-intervals can be compared to
investigations of RR increments -- the popular measure of HRV.
However, the adopted scheme additionally serves as a classification
with respect to the size of increments.  Therefore, due to the network
representation we learn not only what kind of increments dominates
in a signal but also when these events take place.

\section{Results}

\subsection{Group study: healthy versus patients after HTX}

\begin{figure}
\includegraphics[height=19em]{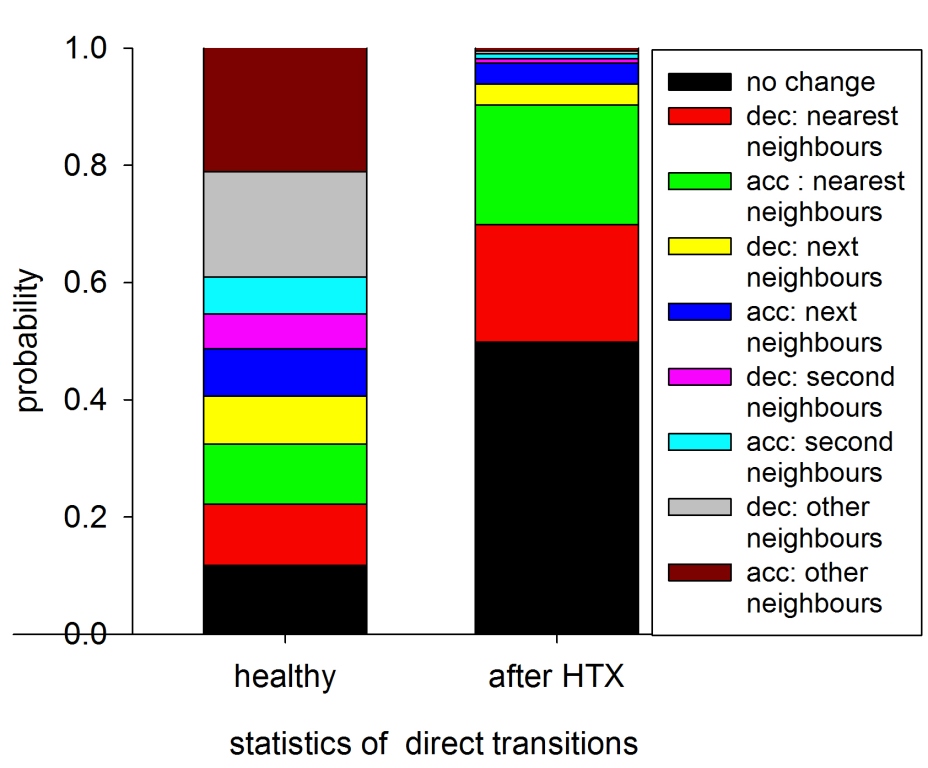}
\caption{Distributions of changes in consecutive RR-intervals for
healthy young people (of age: 18--26, 21 women and 14 men) and
the group of considered patients after HTX. Abbreviations
{\em acc} and {\em dcc} refer to acceleration and deceleration.
For other details, see the main text.}
\label{Fig3}
\end{figure}

The classification of transitions  scheme introduced in Sec. II.B leads to
clear distinction between the heart rhythm of healthy
individuals and those of people after HTX. This is shown in
Fig.~\ref{Fig3}. For example, the transitions to the second and
farther neighbors, in the case of healthy young persons,
occur with probability $0.5$ (upper part of the left bar),
while in  case of people after HTX these events are quite rare.

The majority of transitions in patients after HTX can be described
as no change events. Moreover the decelerating or accelerating
transitions occur rarely. Sequences  of monotonic increases
of length 3, (i.e., sequences in which
$RR_{i} \! < \! RR_{i+1} \! < \! RR_{i+2} \! < \! RR_{i+3}$) occurred,
on average, once in the whole signal. Similar statistics holds for decreases.
Therefore, only monotonic sequences of the length equal to 1 (called one-step mono-transitions) or equal to 2 (two-step mono-transitions) describe  the short-time dynamics of the heart contractions of the HTX patients.

\subsection{Transitions in individuals}
\begin{figure}[h]
\center
\includegraphics[height=14em]{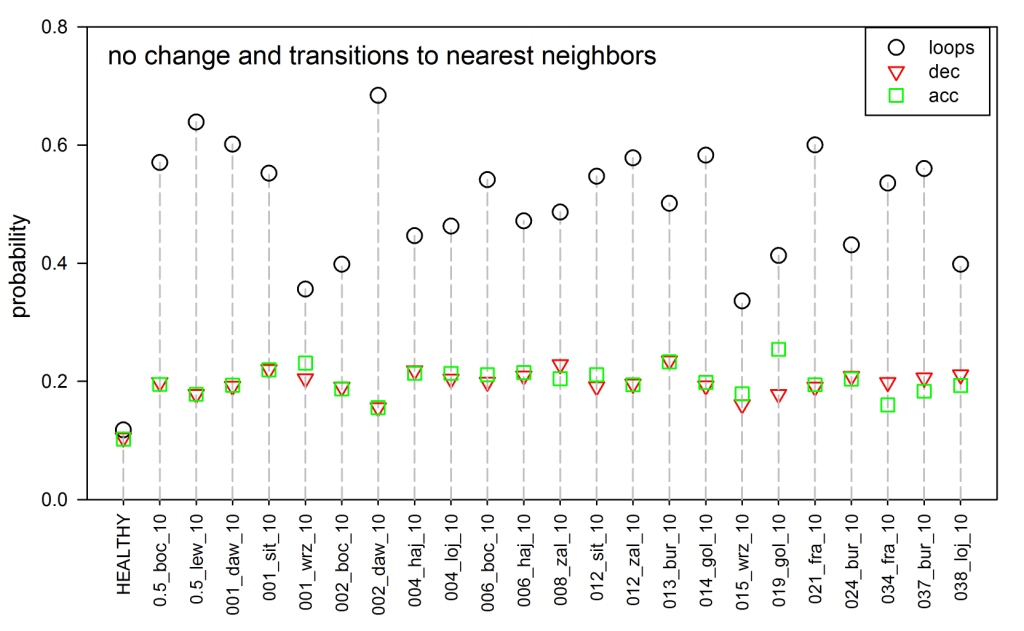}
\includegraphics[height=14em]{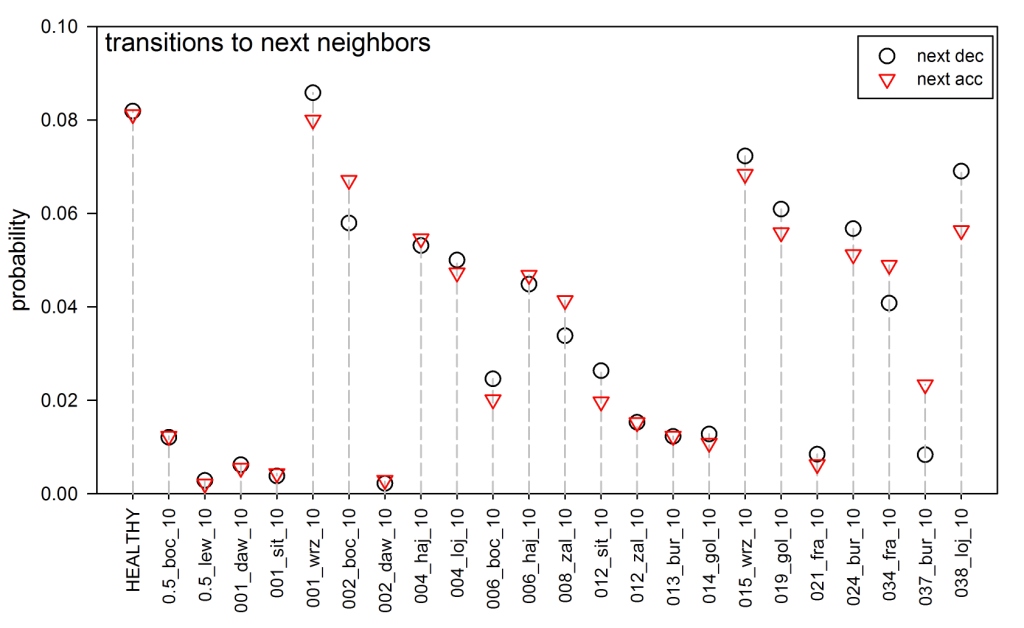}
\caption{Probability to find a given type of transition in a signal
for different patients after HTX: {\it top} -- no change or
transitions to nearest neighbors; { \it bottom} -- transitions
to the next neighbors.}
\label{Fig4}
\end{figure}

Statistics of one-step mono-transitions found for each recording separately are presented in Fig.~\ref{Fig4}. They are compared to the corresponding values calculated for young, healthy young persons (the first entry on horizontal axis).

The first observation from Fig.~\ref{Fig4} top is  that plotted values seem to be independent of the time elapsed after the HTX. As already observed, the no-change or transitions to the nearest neighbor dominate. It means that almost all increments are less then 10 ms. The transitions to the next-neighbors shown in Fig.~\ref{Fig4} bottom (the increments are grater than 10 ms but lower than 20ms) do not provide any regular picture. Moreover also differences between number of  accelerations and
decelerations are statistically not significant. However, concentrating on characteristics for  each patient individually, we get hints  whether the changes are evolving
towards the healthy people characteristics or not.

\begin{figure}
\center
  \includegraphics[height=13em]{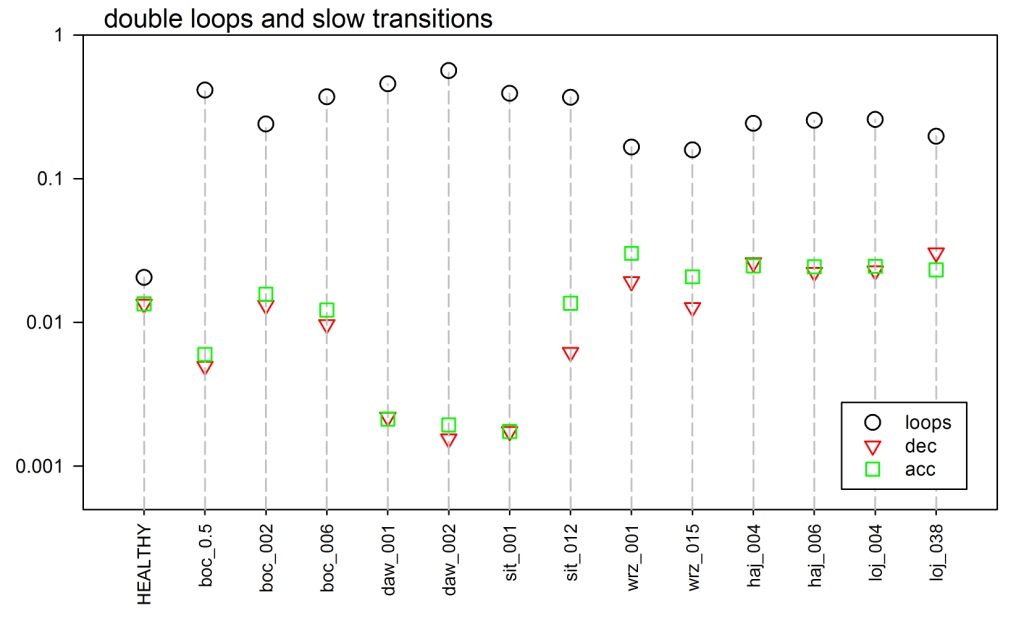}
   \includegraphics[height=13em]{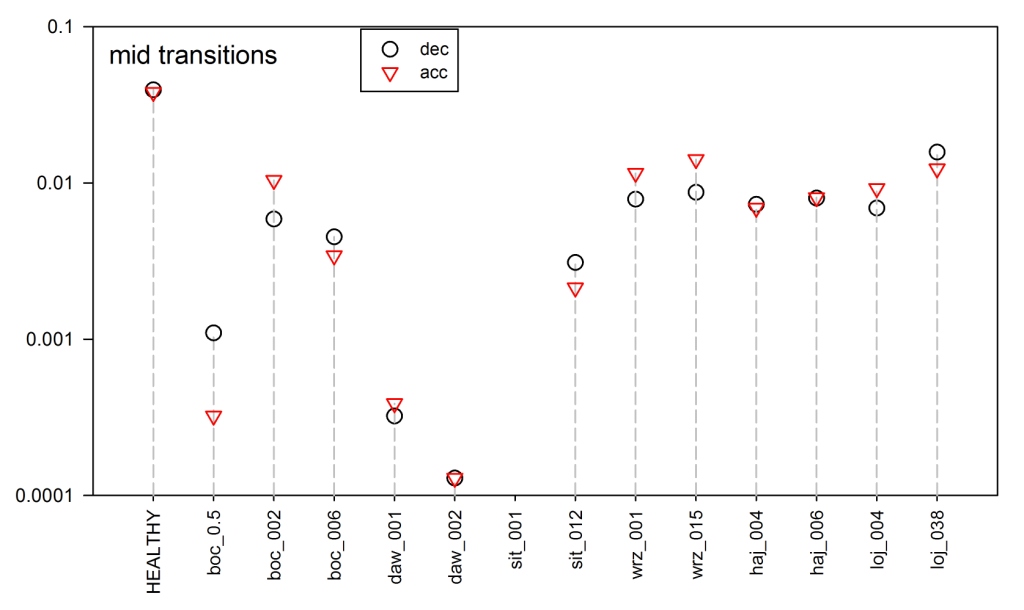}
\caption{Log plots of the probability to observe double loop,
slow ({\it top}) and mid ({\it bottom }) transitions in signals
of patients after HTX. To observe restitution of cardiac control
in a patient, the results are collected by a patient name.
In the case of {\it sit} patient after a month after HTX no mid
transitions occur.}
\label{Fig5}
\end{figure}

The results obtained for the carefully selected  two-step mono-transitions are reported in Fig.~\ref{Fig5}. We classified the monotonic changes  with respect to the total size of the transition. Namely, the consecutive
three RR-intervals $RR_i$, $RR_{i+1}$ and $RR_{i+2}$ are quantified as:\\[2mm]
\hspace*{4mm}double loop:\\
       \hspace*{8mm} $RR_i=RR_{i+1}=RR_{i+2}$; \\
\hspace*{4mm}slow deceleration: \\
       \hspace*{8mm} $RR_i< RR_{i+1} < RR_{i+2}
                     \mbox{~and~}
                     |RR_{i+2} - RR_{i}| = 2\Delta $; \\
\hspace*{4mm}mid deceleration: \\
       \hspace*{8mm} $RR_i \! < \! RR_{i+1} \! < \! RR_{i+2}
                     \mbox{~and~}
                     |RR_{i+2} \! - \! RR_{i}| = 3
                     \mbox{~or~} 4 \Delta $.\\[2mm]
The classification of the corresponding acceleration events goes
with the changed directions of the above  inequalities.

Probabilities of the occurrence of such two-step mono-transitions, shown in Fig.~\ref{Fig5},  are ordered with respect to the patient name, and restricted to recordings from patients being less than about
12 months after the surgery. This ordering helps to track alternations
in the cardiac rhythm in the particular patient. The entry corresponding
to healthy young people is added for comparison.
Note that the plots are in log-scale.

The data presented in Figs~\ref{Fig4} and \ref{Fig5} give a total picture of changes in heart rhythm of a patient. Both analysis: one-step mono-transitions and  two-step mono-transitions  may be useful
in  assessing the progress of patient's recovery, and possibly to
produce an alarming signal that the progress is not satisfactory.

\begin{figure}
    \includegraphics[height=17.2em]{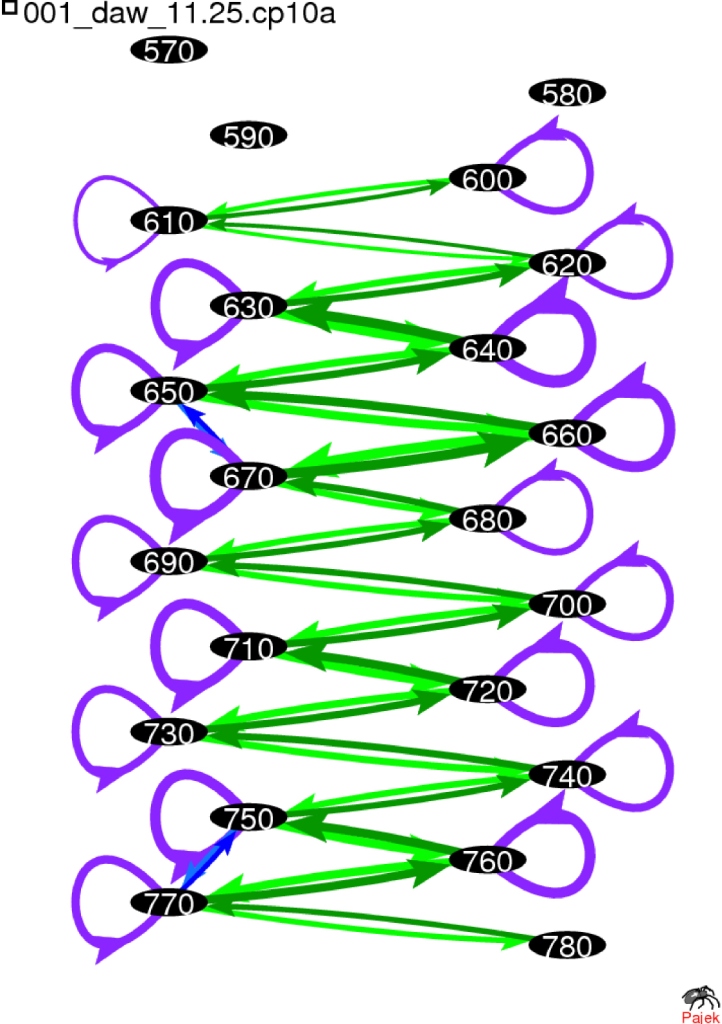}
    \includegraphics[height=17.2em]{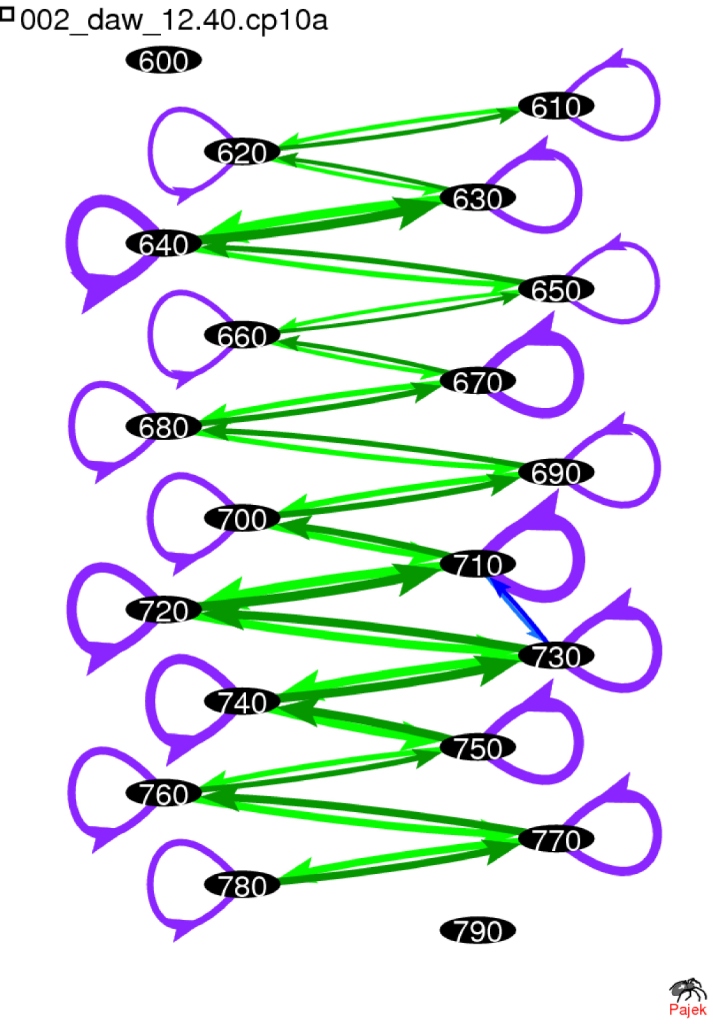}
\caption{Networks for {\it daw} patient obtained from signals
recorded after 1 and 2 months months after HTX.  Transitions
with probability less than 0.1\% are omitted  but all vertices
which  correspond to all recorded RR-intervals are plotted.
The widths of arrows represent  logarithms of counts of the
particular transitions.}
\label{Fig6}
\end{figure}

\begin{figure}
    \includegraphics[height=17.2em]{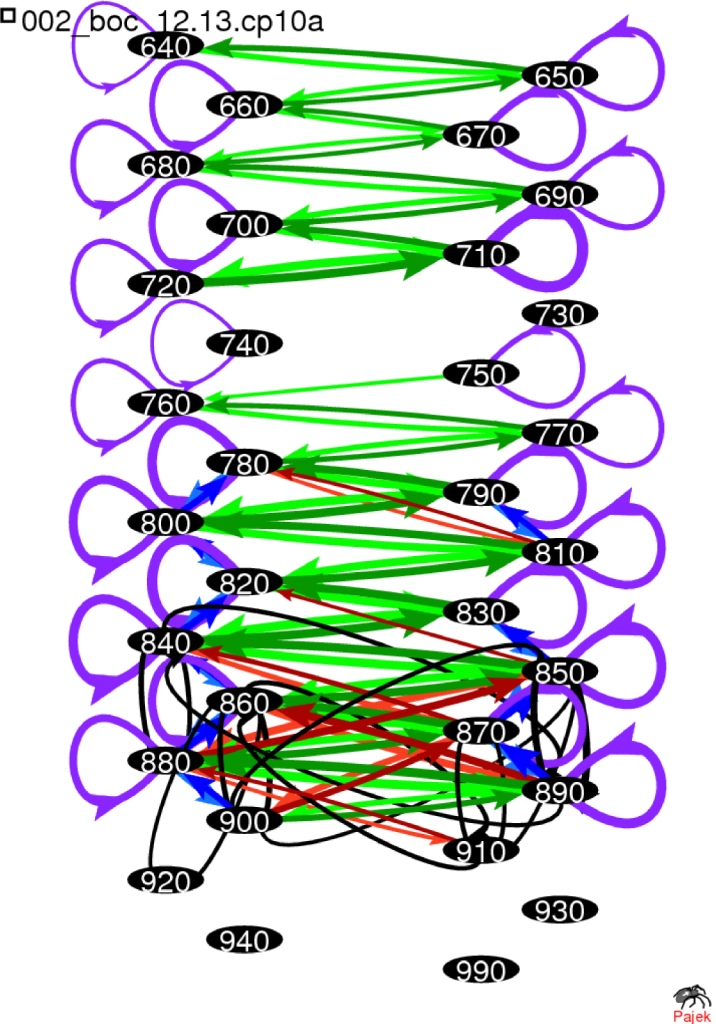}
    \includegraphics[height=17.2em]{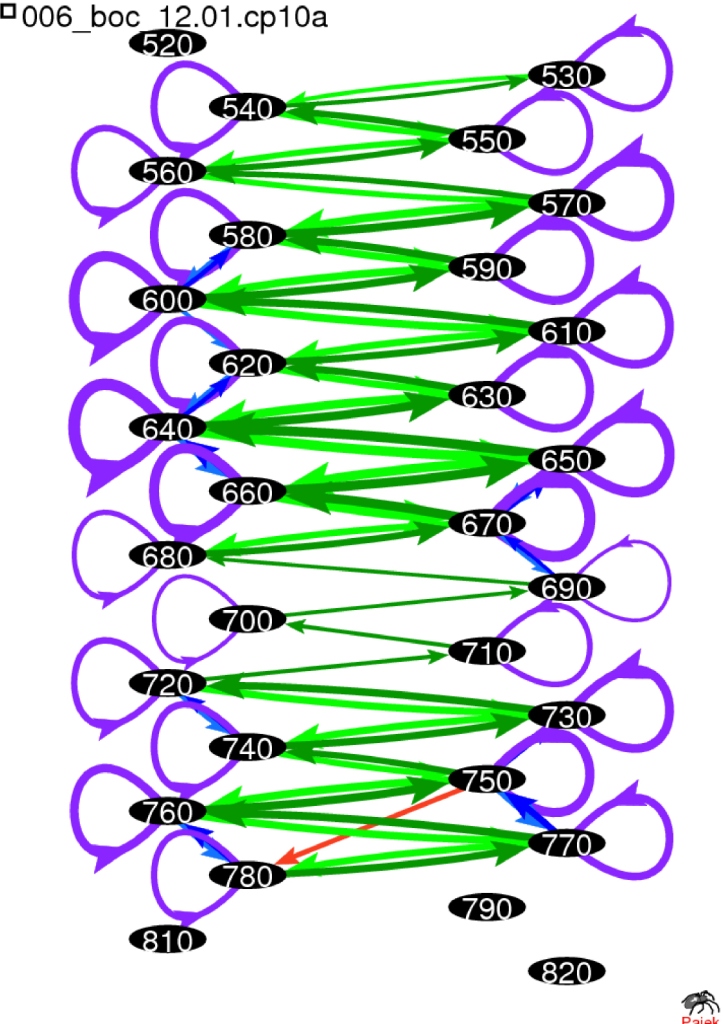}
\caption{Networks for {\it boc} patient obtained from signals
recorded after  2 and 6  months after HTX. }
\label{Fig7}
\end{figure}

For example, Figs~\ref{Fig4} and \ref{Fig5} does not provide the clear picture for the direction of changes of patients {\em daw} and {\em boc}.
Therefore,  via the network presentation we get the eye catching and easily readable additional information on  quality and quantity
of changes. In Figs~\ref{Fig6}, \ref{Fig7} we present network
representation for these  two patients.

The networks of {\it daw} patient (Fig.~\ref{Fig6}) show the
rhythm after 1 and 2 months after surgery. We see that networks
are quite similar, what indicates that the picture is stable.
On the other hand, the networks obtained for {\it boc} patient
(Fig. \ref{Fig7}) show gradual simplification of the cardiac
rhythm structure.  It is worth noting that our predictions based on the analysis of the constructed networks (Figs. \ref{Fig6} and \ref{Fig7})
coincide with the clinical state of these two patients.

\section{Conclusions}

The network of transitions provides yet another way to assess the heart
rhythm. It appears that RR-series leads to the transition network with
the specific shape. Therefore one can classify typical  properties of
these networks, and then construct a measure of heart rhythm changes.
The network representation, first of all, offers a total assessment
of increments between the consecutive RR-intervals by quantifications
and qualification of their values. Additionally, it gives the eye-catching
picture of RR-intervals as a map from which one can read at what
RR-interval and how frequently the particular increase occur.
In this work  we used this approach to observe the restitution of
cardiac control in patients after heart transplantation.

The arguments supporting the decision to perform HTX were
different for each patient. The clinical state of every patient
was specific and, therefore, each time series should be analyzed
individually, and also the progress  in the process of the acceptance
of the graft had to be evaluated  for each patient separately.
This evaluation was attempted due to quantification and classification
of transitions in the phase space of his/her RR-intervals.
We propose to interpret the alternations in the number of  particular
type transitions, here towards corresponding values found in  the healthy
people rhythms,  in the prognosis for individual patient.
Moreover, since sequences of accelerations and decelerations of heart rate
are considered  as a sign of autonomic control \cite{Piskorski2011},
then consecutive intervals with fixed acceleration or deceleration rates
give us additional insights into the activity of the control mechanisms.

\ifCLASSOPTIONcaptionsoff
  \newpage
\fi


\begin{thebibliography}{1}
%
%
\bibitem{Toledo2002}
   R.~Toledo, I.~Pinhas, D.~Aravot, Y.~Almog and  S.~Akselrod
   \emph{t Func\-tio\-nal restitution of cardiac control in heart
   transplant patients}\hskip 1em plus 0.5em minus 0.4em \relax
   Am. J. Physiol. Reg. Integrative Comp.Physiol. 282:R900--R908, 2002.
\bibitem{TASK}
   {\it Heart rate variability. Standards of measurement, physiological
   interpretation, and clinical use}\hskip 1em plus 0.5em minus 0.4em \relax
   Eur.~Heart J.~17:354--81, 1996.
\bibitem{Donner}
   RV.~Donner, Y.~Zou, J.~F.~Donges, N.~Marwan and J.~Kurths
   \emph{ Recurrence networks -a novel paradigm for nonlinear time series
   analysis} \hskip 1em plus 0.5em minus 0.4em \relax
    New J. of Phys. 12: 033025,  2010.
\bibitem{Costa}
   LDaF.~Costa, FA.~Rodrigues, G.~Travieso and PR.~ Villad Boas,
   \emph{t Characterization of comp[lex networks: a survey of measurements}
   \hskip 1em plus 0.5em minus 0.4em \relax Advances in Physics 56:167--242, 2007.
\bibitem{Makowiec2011}
   D.~Makowiec, J.~Wdowczyk-Szulc, M.~\.Zarczyñska-Buchowiecka,
   A.~Ryn\-kiewicz and  M.~Grucha\l a  \emph{t Study heart rate by tools from
   complex networks} \hskip 1em plus 0.5em minus 0.4em \relax
   Acta Phys.~Pol.~B Proc.~Suppl. 4:139-153, 2011
\bibitem{Piskorski2011}
   J.~Piskorski and P.~Guzik,
   \emph{ Structure of heart rate asymmetry: deceleration and acceleration runs}
   \hskip 1em plus 0.5em minus 0.4em \relax Physiol.Meas. 32:1--13, 2011.
\bibitem{pajek}
   \emph{http://pajek.imfm.si/}
%
\end{thebibliography}
\end{document}